\documentclass[aps,prl,twocolumn,superscriptaddress,showpacs,preprintnumbers,longbibliography]{revtex4-1}
\usepackage[colorlinks,bookmarks=false,citecolor=blue,linkcolor=red,urlcolor=blue]{hyperref}
\usepackage{physics}
\input pdfcolor.tex
\usepackage{colortbl,amsthm,txfonts}
\usepackage{verbatim}
\usepackage{graphicx}
\usepackage{epsfig}
\usepackage{dcolumn}
\usepackage{bm}

\usepackage{amsmath,amssymb}

\makeatletter
\renewcommand*\env@matrix[1][*\c@MaxMatrixCols c]{%                                                                     

  \hskip -\arraycolsep
  \let\@ifnextchar\new@ifnextchar
  \array{#1}}
\makeatother
\usepackage{appendix}

\begin{document}
\title{Temperature tuned Fermi surface topology and segmentation in non-centrosymmetric superconductors}
\author{Madhuparna Karmakar}
\email{madhuparna.k@gmail.com}
\affiliation{Centre for Quantum Science and Technology, Chennai Institute of Technology, Chennai-600037, India.}
\affiliation{Department of Physics, Indian Institute of Technology, Madras, Chennai-600036, India.}

\begin{abstract}
  We report the first comprehensive microscopic description of the thermal fluctuations
  tuned Fermi surface characteristics in a non-centrosymmetric superconductor,  in presence of
  an in-plane Zeeman field. Using a non perturbative approach we demonstrate that short range
  fluctuating superconducting pair correlations give rise to segmentation of the Fermi surface with
  direction dependent pair breaking and hotspots for quasiparticle scattering. Further, a fluctuation driven
  change in the Fermi surface topology is realized, characterized by a shift of the corresponding
  Dirac point from the trivial ${\bf k}=0$ to the non trivial ${\bf k} \neq 0$. Our results provide key
  benchmarks for the thermal scales and regimes of thermal stability of the properties
  of these systems, that are important from the applied perspective to spintronic devices.
  Our theoretical estimates are in remarkable qualitative agreement with the recent differential conductance and
  quasiparticle interference measurements on Bi$_{2}$Te$_{3}$/NbSe$_{2}$ hybrid. A generic theoretical
  framework for the finite momentum scattering of quasiparticles and the associated spectroscopic features is
  proposed, which is expected to be applicable to a wide class of superconducting materials.
\end{abstract}

\date{\today}
\maketitle

\textit{Introduction:}
Intriguing superconducting (SC) phases are realized when the constraint of being equal and opposite in
momentum is lifted off the pairing fermions. The key to break this constraint lies in lifting the spin
degeneracy of the fermionic Bloch states, giving rise to spin-split Fermi surfaces (FS) \cite{tedrow1994}.
The corresponding SC state responds to this non trivial FS by allowing the finite momentum ($q$) scattering of
the quasiparticles (QP). Experimental evidences of finite-$q$ scattering include, point contact spectroscopy
showing gapless superconductivity \cite{rybaltchenko1999} and anisotropic FS in rare earth quaternary borocarbides
(RTBC) \cite{baba2008,dreschler2009}, specific heat
and magnetic torque measurements supporting
Fulde-Ferrell-Larkin-Ovchinnikov (FFLO) superconductivity in two-dimensional (2D) organic superconductors
\cite{beyer2012,wosnitza1999,wright2011,mayaffre2014,koutroulakis2016}, scanning tunneling spectroscopy
in favor of in-gap states in magnet-superconductor
hybrid (MSH) \cite{conte2022} and the very recent quasiparticle interference
(QPI) measurements on non-centrosymmetric Bi$_{2}$Te$_{3}$/NbSe$_{2}$ hybrid showing
FS segmentation \cite{jia_science2021}.
%%%%%%%%%%%%%%%%%%%%%%%%%%%%%%%%%%%%%%%%%%%%%%%%%%%%%%%%%%%%%%%%%%%%%%%%%%%%%%%%%%%%%%%%%%%%%%%%%                     
\begin{figure}
\begin{center}
\includegraphics[height=5.0cm,width=8.5cm,angle=0]{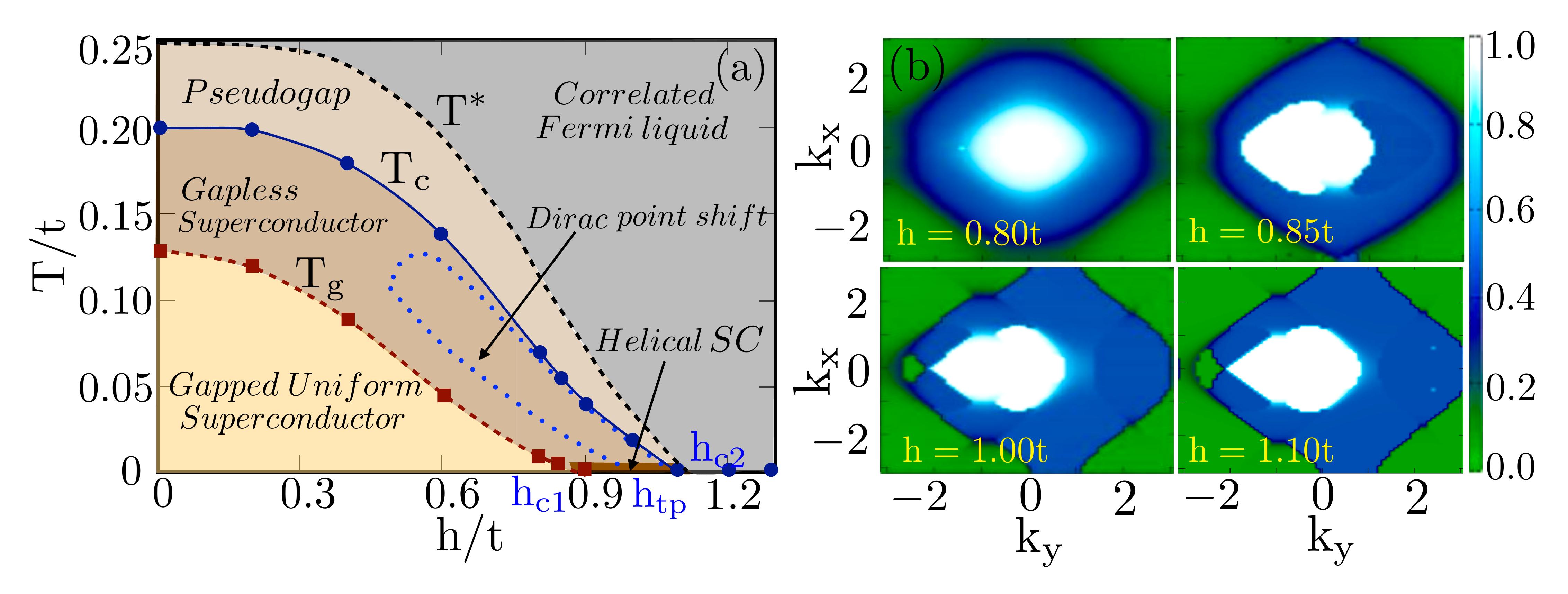}
\caption{(a) Thermal phase diagram in the $h-T$ plane at $\lambda=0.65t$,
showing the thermal scales $T_{c}$, $T_{g}$ and $T^{*}$. The high temperature regime bounded by the dotted curves
correspond to the fluctuation induced change in the FS topology, characterized by the shift in the Dirac point
from ${\bf k}=0$ to ${\bf k} \neq 0$.
(b) Evolution of FS topology with $h$, presented in terms of the fermionic
occupation number $n({\bf k})$. Note the single self intersecting FS at $h=t$}
\label{fig1}
\end{center}
\end{figure}
%%%%%%%%%%%%%%%%%%%%%%%%%%%%%%%%%%%%%%%%%%%%%%%%%%%%%%%%%%%%%%%%%%%%%%%%%%%%%%%%%%%%%%%%%%%%%%%%%

%%%%%%%%%%%%%%%%%%%%%%%%%%%%%%%%%%%%%%%%%%%%%%%%%%%%%%%%%%%%%%%%%%%%%%%%%%%%%%%%%%%%%%%%%%%%%%%%%                     
\begin{figure*}
\begin{center}
\includegraphics[height=6.5cm,width=14.0cm,angle=0]{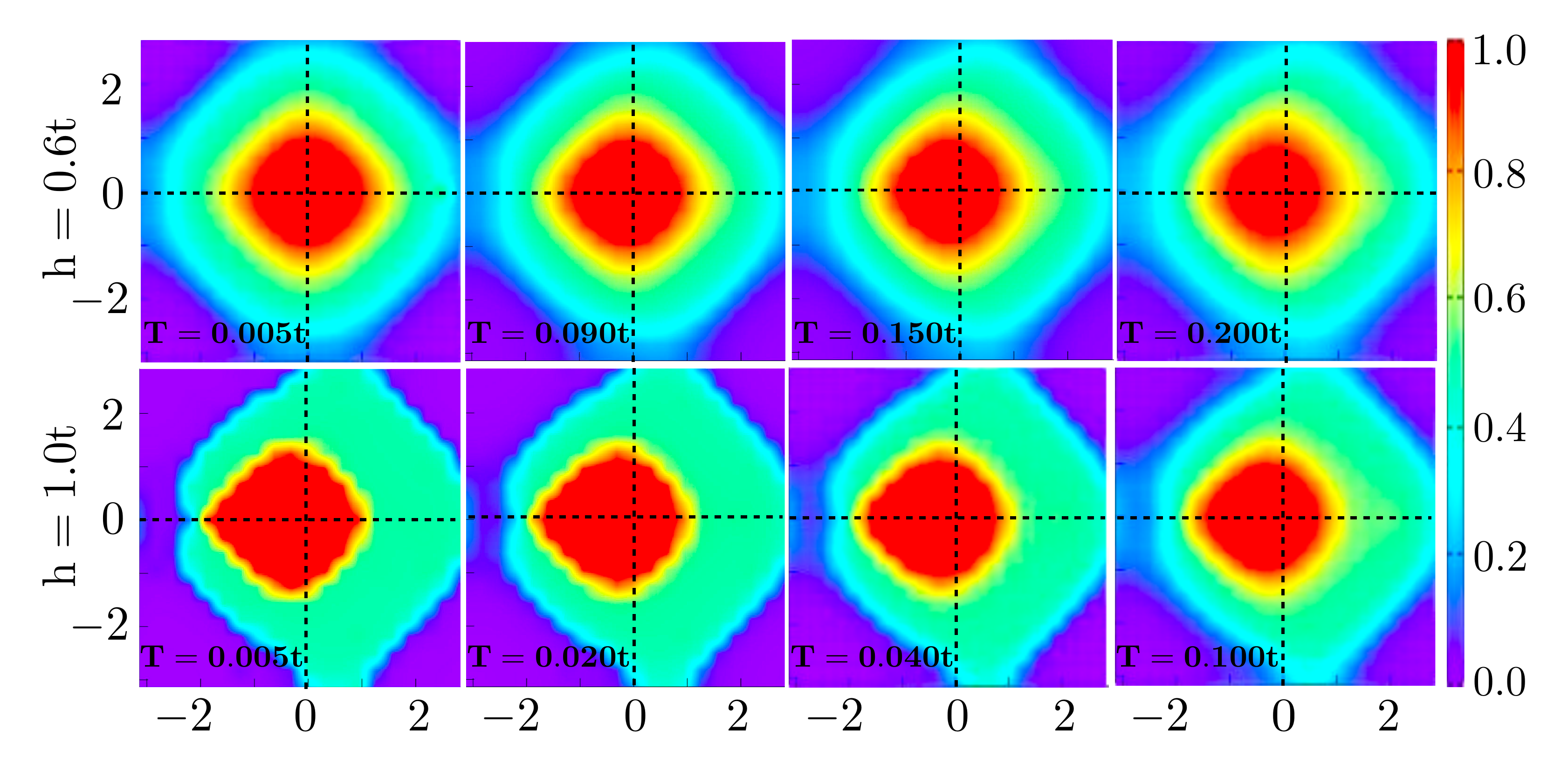}
\caption{Temperature dependence of FS topology at intermediate ($h=0.6t$)
  and strong ($h=t$) Zeeman fields. The dashed lines are guide to the eyes for ${\bf k}=0$.
  Note the self intersecting FS at $T=0.005t$ for $h=t$.}
\label{fig2}
\end{center}
\end{figure*}
%%%%%%%%%%%%%%%%%%%%%%%%%%%%%%%%%%%%%%%%%%%%%%%%%%%%%%%%%%%%%%%%%%%%%%%%%%%%%%%%%%%%%%%%%%%%%%%%%
Non-centrosymmetric superconductors (NCS) and their response to in-plane Zeeman field have recently garnered interest
\cite{kane2008, alicea2010, oreg2010, lutchyn2010, potter2011, mourik2012, shtrikman2012, jia2016, ye2015, saito2015, mak2015,law2016,berg2010,nagaosa2012,fradkin2016,martin2017,fu2018}
owing to the experimental observation of FS segmentation \cite{fu2018,jia_science2021} and SC diode effect
\cite{ono_nature2020, nagaosa2017,yanase_prl2021,nagaosa_njp2021,nagaosa2018,fu2022,jian2022} in these materials.
QPI measurements on Bi$_{2}$Te$_{3}$/NbSe$_{2}$ hybrid showed that the FS corresponding to the
finite-$q$ paired state is segmented, with $q$-dependent isolated parts of the FS
serving as hotspots for QP scattering \cite{jia_science2021}.
In a similar spirit, magnetic field controlled SC diode effect has been observed in non-centrosymmetric
artificial superlattice [Nb/V/Ta]$_{n}$ \cite{ono_nature2020}. Making use of the magnetochiral
anisotropy (MA) of this material, a direction dependent zero resistance to the current flow was
demonstrated \cite{ono_nature2020}. 
The high temperature regime of such systems are shown to be immensely significant from the point
of view of device applications. It was recently demonstrated that the performance of SC tunnel diode made up of
Cu/EuS/Al tunnel junction is robust against thermal fluctuations upto a significantly high operating
temperature, a property that makes it appealing for electronic devices \cite{giazotto2022}.
Similarly, magnetotransport measurements in the SC state of gated MoS$_{2}$ demonstrated that the MA of this
material is strongly dependent on temperature induced SC fluctuations and the ab-initio estimates
of the MA parameter shows a discrepancy of five times w. r. t. the
experimental observations \cite{nagaosa2017}. These experiments suggest that the need of the hour
is to have a theoretical understanding of the impact of thermal fluctuations on NCS.

Theoretical attempts to analyze the observations on NCSs have remained largely restricted
to mean field theory (MFT) and  perturbative approaches,  which though are well 
suited to access the ground state properties, fail to capture the fluctuation effects
\cite{fu_pnas2021,akbari_prr2022,akbari2022,fu_natcom2021,fu2018,yanase_prl2021,fu2022}.
Access to thermal physics requires a beyond MFT approach to the problem and the complex parameter
space with at least five competing energy scales viz. pairing interaction, spin-orbit coupling (SOC),
Zeeman field, electron hopping and temperature, poses a computationally difficult problem.

In this letter we take a look at the 2D-NCS in the light of a non
perturbative numerical approach that retains the spatial fluctuations of the SC
pairing field at all orders. In particular, we investigate the finite temperature physics
of this system in the space of competing pairing interaction ($U$), SOC ($\lambda$) and
in-plane Zeeman field ($h$), while retaining the temperature ($T$) induced SC
fluctuations. Our results provide estimates for the regime of thermal stability of the
magnetochiral properties of these materials, which is important for spintronic devices.

Based on thermodynamic and spectroscopic signatures our key observations are as follows:
$(i)$ the system comprises of Zeeman field tuned two quantum critical points (QCP), $h_{c1}$ and
$h_{c2}$,  corresponding to a first order phase transition, with a shift in the Dirac point to
${\bf k} \neq 0$, between the uniform and the helical SC (discussed later) phases,  and a second
order transition between the helical SC and a correlated Fermi liquid (CFL) phase, respectively.
$(ii)$ A topological transition (TT) of the FS takes place at $h=h_{tp}$, accompanied by a
crossover between the interband and intraband paired helical SC phases. We note that the QCP $h_{c1}$
is not tied to $h_{tp}$ and no symmetry breaking takes place across the TT.  
$(iii)$ Thermal fluctuations alter the FS topology and a temperature controlled shift of
the Dirac point from the original ${\bf k}=0$ to ${\bf k}\neq 0$ is realized
for $T \neq 0$. $(iv)$ FS segmentation takes place in the helical SC state, governed by thermal
fluctuations generated hotspots for QP scattering. $(v)$ Our results are in remarkable qualitative
agreement with the experimental observations on Bi$_{2}$Te$_{3}$/NbSe$_{2}$ hybrids \cite{jia_science2021}.
The Zeeman field tuned evolution of the in-gap states as observed in differential conductance measurements
as well as the FS segmentation observed via QPI measurements on Bi$_{2}$Te$_{3}$/NbSe$_{2}$ are well
captured by our theoretical framework. Our results at $T \neq 0$ are likely to trigger important high
temperature experiments to realize thermally controlled MA in NCS. 

\textit{Model and observable:}
Our starting Hamiltonian is the 2D attractive Hubbard model with
Rashba spin-orbit coupling (RSOC) and in-plane Zeeman field \cite{zhang_soc2014,wu2013,agterberg_rev2017},
\begin{eqnarray}
H & = & -t\sum_{\langle ij\rangle, \sigma}(c^{\dagger}_{i, \sigma}c_{j, \sigma} + h. c.) 
-\mu \sum_{i, \sigma}\hat n_{i, \sigma} - \vert U \vert\sum_{i}\hat n_{i, \uparrow}\hat n_{i, \downarrow}
\nonumber \\ && 
+ \lambda\sum_{ij, \sigma \sigma^{\prime}}(c^{\dagger}_{i, \sigma}(i\hat \sigma_{y})_{\sigma, \sigma^{\prime}}
c_{j, \sigma^{\prime}} + c^{\dagger}_{i, \sigma}(-i\hat \sigma_{x})_{\sigma, \sigma^{\prime}}c_{j,\sigma^{\prime}})
\nonumber \\ && + h\sum_{i}(c^{\dagger}_{i, \uparrow}c_{i, \downarrow} + c^{\dagger}_{i, \downarrow}
c_{i, \uparrow})
\label{Eqn1}
\end{eqnarray}
\noindent where, $t=1$ is the hopping amplitude between the nearest neighbors on a square lattice and
sets the reference energy scale of the problem, $\lambda$ is the magnitude of RSOC.  $\vert U\vert > 0$
is the on-site attractive interaction between the fermions and $\mu$ is the global chemical potential which
fixes the electron density. The in-plane Zeeman field ($h$)  
is applied along the $x$-axis; $\hat \sigma_{x}$ and $\hat \sigma_{y}$ are the Pauli matrices.

The non-interacting ($U = 0$) energy dispersion of Eq.\ref{Eqn1} reads as, $E_{\bf k}^{\eta}=\xi_{\bf k} \pm \sqrt{h^{2}-2ih\lambda \sin k_{y}-\lambda^{2}(\sin^{2}k_{x}+\sin^{2}k_{y})}$, where,  $\xi_{\bf k}=-2t(\cos k_{x}a + \cos k_{y}a)-\mu$ is the kinetic energy
contribution, with lattice spacing $a$. $E_{\bf k}^{\eta}$ corresponds to the helicity
bands labeled by the helicity index $\eta=\pm$. In the limit of $h=0$,  the spectra comprises of four
dispersion branches and with $h \neq 0$ each branch splits into two \cite{agterberg_rev2017}. 
We make the interacting ($U \neq 0$) model numerically tractable via Hubbard Stratonovich (HS)
decomposition of the quartic term \cite{hs1,hs2}. This introduces a complex scalar
bosonic auxiliary field $\Delta_{i}(\tau)=\vert \Delta_{i}(\tau)\vert e^{i\theta_{i}(\tau)}$ at each site,
which couples to the SC pairing channel.
%%%%%%%%%%%%%%%%%%%%%%%%%%%%%%%%%%%%%%%%%%%%%%%%%%%%%%%%%%%%%%%%%%%%%%%%%%%%%%%%%%%%%%%%%%%%%%%%%                     
\begin{figure}
\begin{center}
\includegraphics[height=8.5cm,width=8.5cm,angle=0]{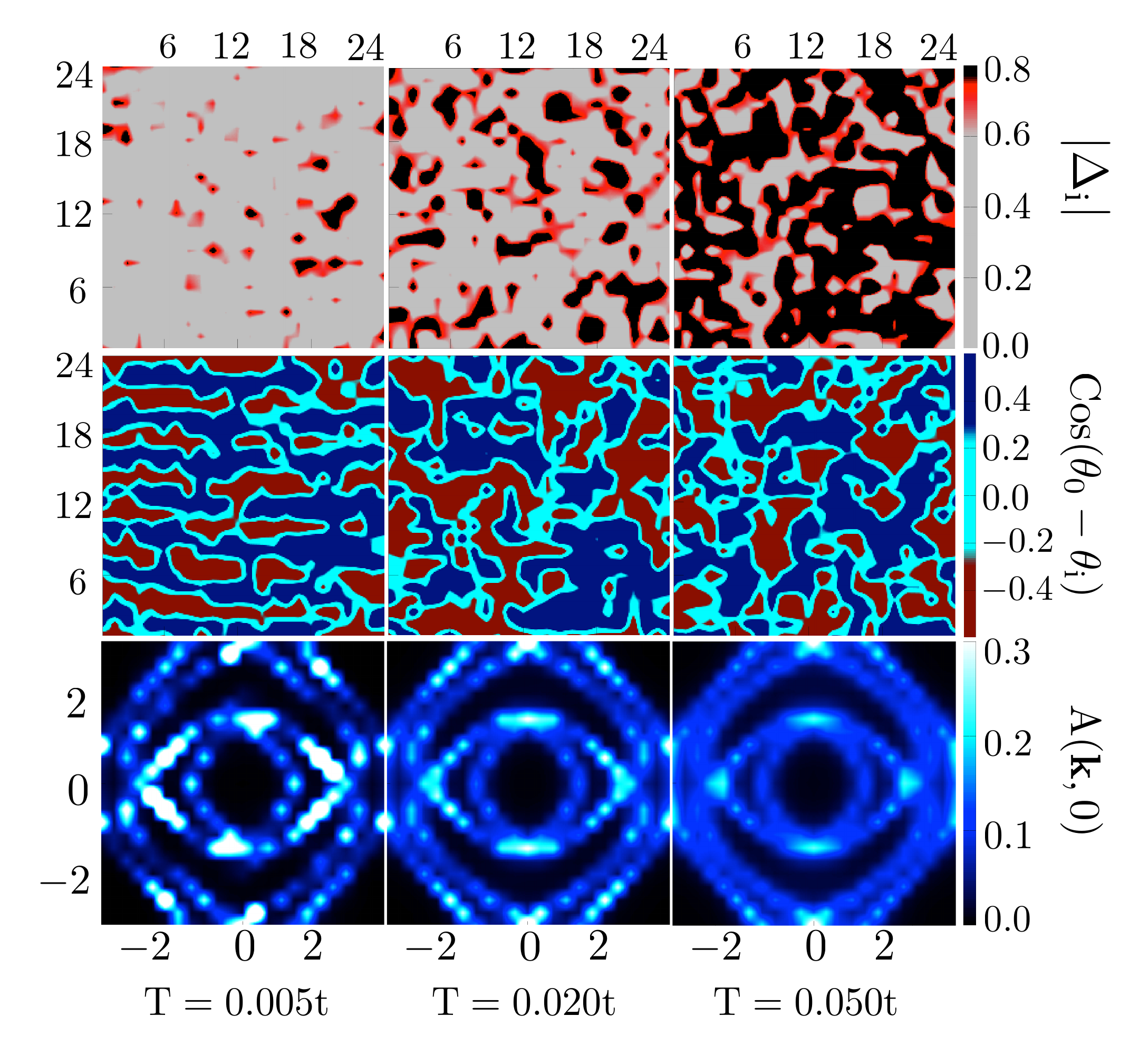}
\caption{Real space maps (in the $xy$-plane) corresponding to the SC pairing field amplitude, $\vert \Delta_{i}\vert$
(top row) and SC phase coherence, $\cos(\theta_{0}-\theta_{i})$ (middle row), showing the
thermal evolution of the helical SC state at $h=t$. The bottom row maps out the segmented FS in terms of
the low energy spectral weight distribution, $A({\bf k}, 0)$ (in the $k_{x}k_{y}$-plane). Thermal
fluctuations lead to fragmentation of the underlying SC state and progressive accumulation of spectral
weight such that the FS isotropy is restored at $T \sim 0.05t$.}
\label{fig3}
\end{center}
\end{figure}
%%%%%%%%%%%%%%%%%%%%%%%%%%%%%%%%%%%%%%%%%%%%%%%%%%%%%%%%%%%%%%%%%%%%%%%%%%%%%%%%%%%%%%%%%%%%%%%%%

We address this problem using static path approximated (SPA)
quantum Monte Carlo technique, wherein we treat $\Delta_{i}$ as classical by retaining its complete spatial
fluctuations but taking into account only the $\Omega_{n}=0$ Matsubara mode in frequency
(i. e. $\Delta_{i}(\tau) \rightarrow \Delta_{i}$) \cite{mpk2016,mpk2018,mpk_jpcm2020,mpk_lieb2020}
(see supplementary materials (SM)). 
The equilibrium configurations of $\{\Delta_{i}\}$ (both for the uniform ${\bf q}=0$ and the
helical ${\bf q} \neq 0$ states) are generated via Monte Carlo (MC) simulation and the 
different fermionic correlators are computed on these equilibrium configurations.  
We verify our MC simulation results at the ground state using an alternate scheme of
Bogoliubov-de-Gennes mean field theory (BdG-MFT).
%%%%%%%%%%%%%%%%%%%%%%%%%%%%%%%%%%%%%%%%%%%%%%%%%%%%%%%%%%%%%%%%%%%%%%%%%%%%%%%%%%%%%%%%%%%
\begin{figure*}
\begin{center}
\includegraphics[height=8cm,width=15cm,angle=0]{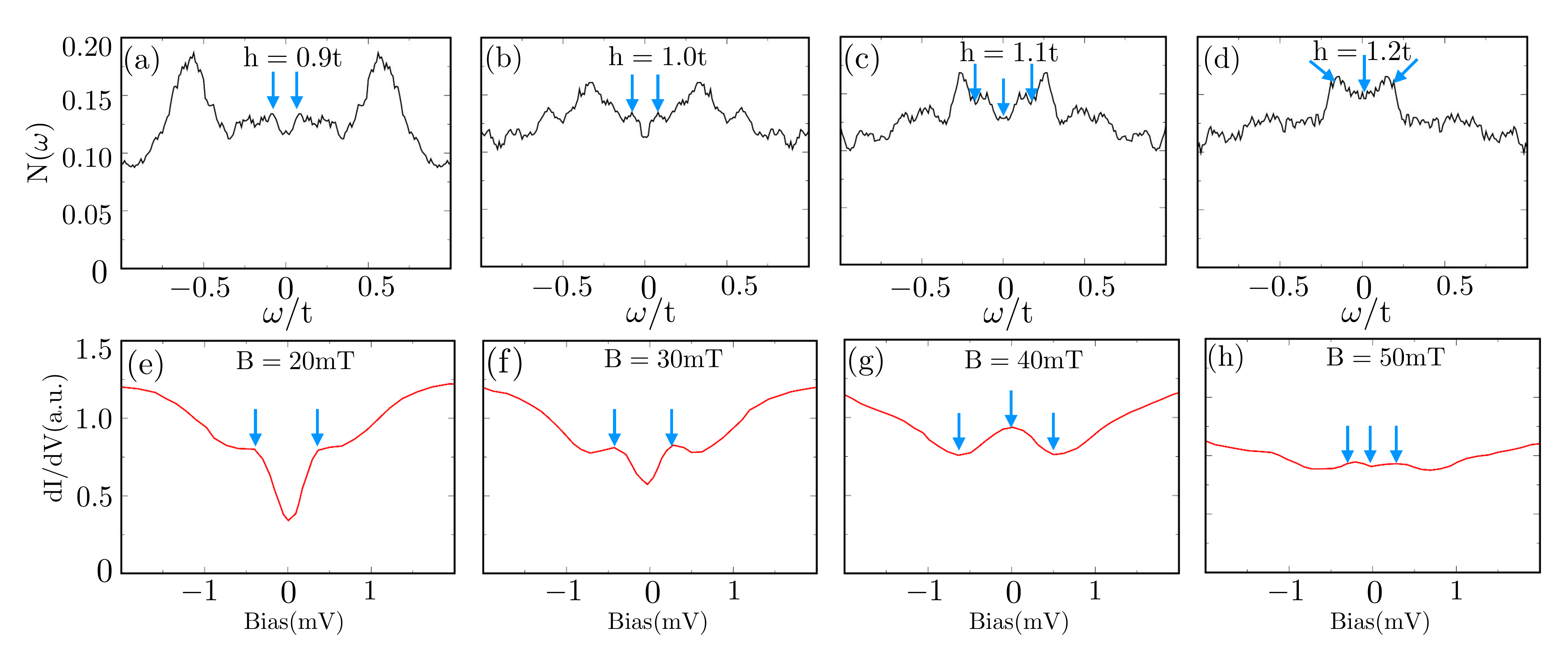}
\caption{Comparison of the evolution of the in-gap states with Zeeman field in
  the helical SC phase with the experimental observations on Bi$_{2}$Te$_{3}$/NbSe$_{2}$
  hybrid \cite{jia_science2021}. The top panels correspond to the theoretically computed single
  particle DOS as function of the Zeeman field and the bottom panels show the results obtained
  via differential conductance measurement in Bi$_{2}$Te$_{3}$/NbSe$_{2}$ hybrid, as function of the
  applied magnetic field $B || \Gamma-M$ \cite{jia_science2021}. The arrows indicate the in-gap states.}
\label{fig4}
\end{center}
\end{figure*}
%%%%%%%%%%%%%%%%%%%%%%%%%%%%%%%%%%%%%%%%%%%%%%%%%%%%%%%%%%%%%%%%%%%%%%%%%%%%%%%%%%%%%%%%%%%

We characterize the system based on the following indicators, $(i)$ SC pairing field structure factor, $S({\bf q})$,
quantifying the (quasi) long range SC phase coherence;
$(ii)$ single particle density of states (DOS), $N(\omega)$ and the corresponding spectral gap
$E_{g}$ at the Fermi level (zero energy); $(iii)$ low energy spectral weight distribution,
$A({\bf k}, 0)$; $(iv)$ spectral function, $A({\bf k}, \omega)$ and $(v)$ momentum resolved fermionic
occupation number, $n({\bf k})$ (see SM). We work in the grand canonical ensemble with  $\mu=-t$ corresponding
to an electron filling of $n \approx 0.75$,  the pairing interaction is $\vert U\vert = 4t$. The simulations
are carried out on a system size of $L=24$ (for MC) and of $L=70$ (for BdG-MFT), unless specified otherwise. 

\textit{Phase diagram and thermal scales:} Fig.\ref{fig1}(a) constitutes the primary result of this
work, wherein we show the thermal phase diagram of the 2D-NCS in the 
$h-T$ plane, for a selected RSOC of $\lambda=0.65t$. The QCPs at $h_{c1}\approx 0.85t$ and
$h_{c2} \approx 1.1t$, correspond to a first order transition
between gapped uniform (${\bf q} = 0$) and gapless helical (${\bf q} \neq 0$) SC phases and a
second order transition between the helical SC and CFL, respectively.
Within the helical SC phase the TT of the FS at $h_{tp} \sim t$ is associated  
with a crossover between inter and intraband SC pairing. For $h > h_{c2}$,  the SC correlations
are lost. 

The evolution of the FS topology across the QCPs is shown in Fig.\ref{fig1}(b) in terms of the fermionic
occupation number $n({\bf k})$ at selected $h$. $h=0.80t$ is the representative of the $0 < h \le h_{c1}$
regime with uniform SC state. The corresponding concentric FSs are symmetric about the Dirac point
at ${\bf k}=0$. $h=0.85t$ represents the helical SC state in the regime $h_{c1} < h \le h_{tp}$
with the Dirac point shifted to ${\bf k} \neq 0$. At $h_{tp} \sim t$,  the FSs cross each other giving
rise to a {\it single self intersecting} FS akin to the Limacon of Pascal \cite{fu_pnas2021}. No quantum
symmetry is broken across $h_{tp}$ and this transition can't be detected
via the routine spectroscopic and transport probes used to detect SC phase transitions.
Further, increase in $h$ decouples the FSs and pushes them
apart, as shown at $h=1.1t$. Note that the Dirac point is at ${\bf k}\neq 0$ for the
decoupled FSs and even at stronger $h$ when the
SC correlations are lost. The CFL containing FS with Dirac points at ${\bf k} \neq 0$ is thus
topologically different from the polarized Fermi liquid realized in the high magnetic field
regime of FFLO \cite{mpk2016} or the normal state of conventional and uniform superconductors. 
%%%%%%%%%%%%%%%%%%%%%%%%%%%%%%%%%%%%%%%%%%%%%%%%%%%%%%%%%%%%%%%%%%%%%%%%%%%%%%%%%%%%%%%%%%%%%%%%%
\begin{figure*}
\begin{center}
\includegraphics[height=11cm,width=16.5cm,angle=0]{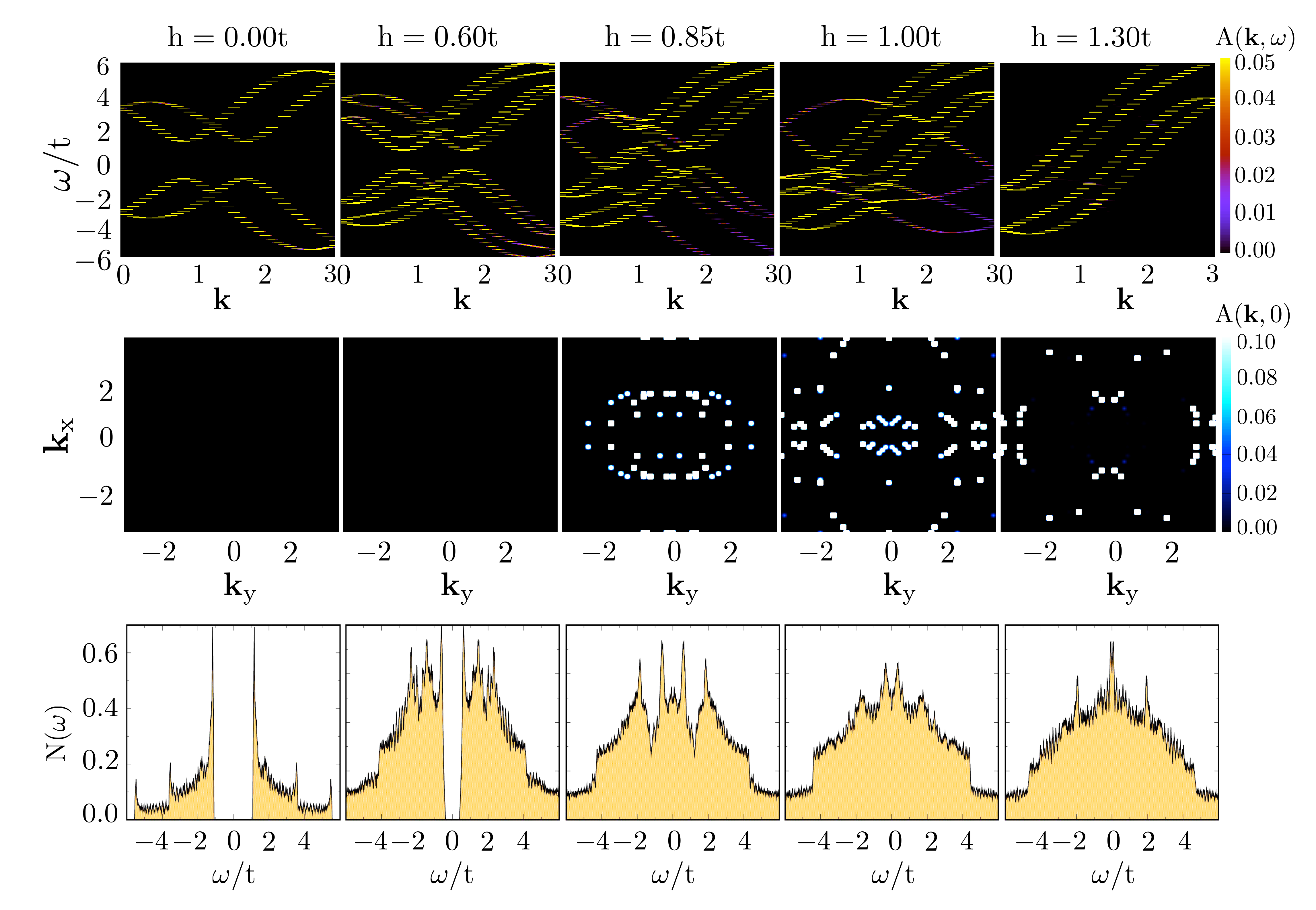}
\caption{Spectroscopic properties, (i) spectral function, $A({\bf k}, \omega)$ (top row),
(ii) low energy spectral weight distribution, $A({\bf k}, 0)$ (middle row) and
(iii) single particle DOS, $N(\omega)$ (bottom row), at the ground state as function of $h$,
determined using BdG-MFT for $L=70$. For $h=0$ uniform interband superconductivity is realized between the
helicity bands, giving rise to a robust zero energy SC gap. The corresponding $A({\bf k}, 0)$ is featureless
and $N(\omega)$ shows prominent gap edge singularities. $h\neq 0$
splits the helicity bands, allows for symmetric finite energy intraband SC pairing and opens up the corresponding
shadow gaps. For $h > h_{c1}$ the system is in the helical SC state and the multi-branched dispersion gives
rise to gapless spectra. The corresponding $N(\omega)$ has
finite spectral weight at the Fermi level and in-gap states. The FS is segmented, as observed via $A({\bf k}, 0)$,
with isolated hotspots for quasiparticle scattering.
The strong $h$ regime ($h > h_{c2}$) corresponds to the CFL phase with anisotropic FS akin to a magnetic metal.}
\label{fig5}
\end{center}
\end{figure*}
%%%%%%%%%%%%%%%%%%%%%%%%%%%%%%%%%%%%%%%%%%%%%%%%%%%%%%%%%%%%%%%%%%%%%%%%%%%%%%%%%%%%%%%%%%%%%%%%%

The finite temperature phases are characterized by three thermal scales, $T_{c}$, $T_{g}$ and
$T^{*}$, corresponding to the SC transition via the loss of global SC phase coherence, collapse
of the zero energy SC spectral gap at the Fermi level and crossover between the pseudogap (PG)
and CFL regimes, respectively. The thermal transition temperature
$T_{c}$ is determined based on the temperature dependence of $S({\bf q})$, while the crossover
scales $T_{g}$ and $T^{*}$ are determined based on the thermal
evolution of $N(\omega)$ at the Fermi level (see SM). The high temperature PG
phase contains short range SC pair correlations, which show up as phase decohered SC islands.
The regime is smoothly connected to the CFL via gradual degradation of the short range
SC pair correlations.

\textit{Temperature tuned FS topology:} Thermal fluctuations induced short range SC
correlations alter the FS topology over the regime enveloped by the dotted curves, in Fig.\ref{fig1}(a).
We understand this behavior based on Fig.\ref{fig2}, which shows the thermal evolution
of the FS at selected Zeeman field of $h=0.6t$ and $h=t$. At $h=0.6t$, the low
temperature phase ($T \le 0.04t$) corresponds to a gapped uniform superconductor, with the Dirac point
at ${\bf k}=0$, followed by a gapless SC regime $0.04t < T \le 0.15t$.
Within the gapless phase, temperature alters the FS topology by progressively shifting the Dirac point
from ${\bf k} = 0$ to ${\bf k}\neq 0$ over the regime $0.09t < T \le 0.15t$,
indicating the dominance of fluctuation induced helical SC correlations.
The FS undergoes fluctuation induced broadening in the PG regime,  $T > T_{c}=0.15t$

At $h=t$, the system is in the helical SC state with a single self intersecting FS, as shown
at $T=0.005t$.  Temperature decouples the FSs and progressively shifts the smaller
FS away from the edge of the larger, over the regime $0 < T \le 0.02t$. This thermal fluctuations
driven transition of the FS topology is accompanied by the crossover between intra and
interband helical SC correlations. Loss of local SC correlations for $T > 0.04t$ is indicated
by the broadening of the FS.

\textit{Helical superconductivity and Fermi surface segmentation:} We next analyze the helical
SC state in terms of its real and momentum space characteristics. Fig.\ref{fig3} shows the temperature
dependence of the SC pairing field amplitude ($\vert \Delta_{i}\vert$) (top row) and phase coherence
($\cos(\theta_{0}-\theta_{i})$) (middle row). The low temperature helical SC phase is characterized by a spatially
uniform SC amplitude and a one dimensional (1D) modulated phase coherence. Fluctuations progressively
destroys the global SC order via loss of (quasi) long range phase coherence. The state undergoes spatial
fragmentation into phase decohered islands with large local pairing field amplitude, corresponding
to the PG phase. 

The bottom row of Fig.\ref{fig3} shows the segmentation of the FS, mapped out in terms of the
low energy spectral weight distribution $A({\bf k}, 0)$. Finite-$q$ pairing leads to direction
dependent pair breaking, such that only parts of the gapless FS are available for the scattering
of the QPs. Temperature leads to the accumulation of spectral weight such that the
FS isotropy is restored for $T \ge 0.05t$. Our results at low temperatures, shown in Fig.\ref{fig3}
are in remarkable agreement with the experimental observations on Bi$_{2}$Te$_{3}$/NbSe$_{2}$
hybrid \cite{jia_science2021}. QPI measurements on this NCS showed FS segmentation
with $q$-dependent QP scattering. The corresponding real space behavior, as established
via differential conductance maps show 1D standing wave modulations \cite{jia_science2021}, akin
to our results in Fig.\ref{fig3}.

\textit{Comparison with experiment and finite energy pairing:} We now compare in Fig.\ref{fig4}, our
low temperature single
particle DOS ($N(\omega)$) at different Zeeman fields with the differential conductance measurement on
Bi$_{2}$Te$_{3}$/NbSe$_{2}$ hybrid \cite{jia_science2021}. In consonance with the experimental
observation, the $N(\omega)$ exhibits Zeeman field tuned evolution of the in-gap states.
We understand this as follows: owing to the finite-$q$ pairing the $\vert {\bf k}_{\uparrow}\rangle$
state not just connects with $\vert {\bf -k}_{\downarrow}\rangle$, but with
$\vert {\bf -k+q}_{\downarrow} \rangle$ and $\vert {\bf -k-q}_{\downarrow} \rangle$ states as well.
The resulting dispersion spectra contains multiple branches and gives rise to additional van Hove
singularities, which shows up as in-gap states in $N(\omega)$.

We understand this physics using Fig.\ref{fig5} where we show the systematic evolution of
the low temperature spectroscopic properties of this system across the QCPs.
At $h=0$, the dispersion spectra of the $s$-wave NCS contains four branches, as shown
via the $A({\bf k}, \omega)$ map. The uniform SC state is interband paired with fermions
belonging to the two helicity bands pairing up. The corresponding spectra is gapped at the zero energy
with sharp van Hove singularities at the gap edges, indicting the (quasi) long
range SC phase coherence. Zeeman field ($h=0.6t$) splits the dispersion branches, the spectral gap
at the zero energy arising from the uniform interband pairing is suppressed in magnitude. 
Additional finite energy ``shadow gaps'' open up, which are symmetrically located at $\omega \approx \pm U/2$.
These shadow gaps which are the replicas of the zero energy gap and recently discussed
in the context of Ising superconductors \cite{belzig2021}, arises out of the intraband uniform
pairing in the individual helicity bands. Thus, the intermediate $0 < h \le h_{c1}$ regime
contains SC state with admixture of uniform inter and intraband pairing. 
In this regime the multiple branches of the dispersion spectra arises out of the interplay of SOC and
in-plane Zeeman field, and doesn't involve any finite-$q$ scattering of the QPs.
Expectedly, the SC pairing continues to be between the $\vert {\bf k}_{\uparrow} \rangle$ and the
$\vert {\bf -k}_{\downarrow} \rangle$ states and there are no FS segmentation and in-gap states
either in the zero or in the finite energy SC gaps.  

The helical SC phase ($h=0.85t$) with finite-$q$ pairing contains multiple dispersion
branches which connects a larger set of states. Some of these dispersion
branches crosses the Fermi level, giving rise to gapless superconductivity.
SC pairing is interband in the regime $h_{c1} < h \le h_{tp}$ and the shadow gaps
are strongly suppressed. Note that a recent MFT study have discussed the possibility of
finite energy finite momentum pairing in Fulde-Ferrell (FF) superconductors, in the absence of RSOC
\cite{black_schraffer2022}. 
The FS of the helical SC state is segmented with direction dependent pair breaking, as shown via 
($A({\bf k}, 0)$). The crossing of the dispersion branches (as observed in $A({\bf k}, \omega)$) give
rise to the additional van Hove singularities and the associated in-gap states, as observed in the corresponding
$N(\omega)$. At $h=h_{tp}=t$, a single self intersecting FS is realized which is reflected
in $A({\bf k}, \omega)$ as self intersecting dispersion branches. The corresponding
SC state undergoes topological transition from the inter to the intraband pairing
and continues to be an intraband helical superconductor over the regime $h_{tp} < h \le h_{c2}$.
SC correlations are lost at $h=1.3t$ and the dispersion spectra is akin to that of a magnetic metal
with anisotropic FS.

\textit{Discussion and conclusions:} Segmentation of FS and multibranched dispersion are 
generic properties associated with finite-$q$ scattering of QPs. The helical SC state
constitutes one of the examples, while the others include FFLO
\cite{torma2007,batrouni2012,torma2013,torma2014,mpk2016,mpk_bookchapter,beyer2012,wosnitza1999,wright2011,mayaffre2014,koutroulakis2016},
MSH \cite{mirlin2019,loss2016,conte2022}, magnetic superconductors
such as, RTBC \cite{baba2008,dreschler2009,kontani2004,mpk_mag_sc,mpk_mag_sc_finite,buzdin_rmp2005} etc.
Our analysis of the SC state based on Fig.\ref{fig5} is generic and is applicable to other systems
as well. Gapless SC phase with anisotropic (nodal) FS has been experimentally observed in YNi$_{2}$B$_{2}$C
and LuNi$_{2}$B$_{2}$C \cite{boaknin2001,watanabe2004,baba2010,yang2000,oguchi2000}, and magnetic
fluctuations were proposed to play a key role in defining the FS and SC gap architecture in these
\cite{kontani2004} and related class of materials \cite{mpk_mag_sc,mpk_mag_sc_finite}.

In a similar spirit, Nuclear magnetic resonance (NMR) measurements on 2D organic superconductor
$\beta^{\prime \prime}$-(BEDT-TTF)$_{2}$SF$_{5}$CH$_{2}$CF$_{2}$SO$_{3}$ (BEDT-TTF) showed that
the high magnetic field low temperature regime hosts FFLO state with 1D modulated SC order
\cite{koutroulakis2016}. de Hass-van Alphen measurements and angle dependent magnetoresistance
oscillations were utilized to map out the highly anisotropic 2D FS of BEDT-TTF \cite{wosnitza1999}.
Theoretically, a non perturbative beyond MFT study of the FFLO phase in an isotropic $s$-wave superconductor
established FS segmentation, in terms of the spectroscopic signatures \cite{mpk_bookchapter}.
Thus, irrespective of its origin, finite-$q$ scattering of QPs bring out similar physics in
widely different classes of SC systems.

In conclusion, based on a non perturbative numerical approach we have investigated the spin split
non-centrosymmetric superconductor in presence of an in-plane Zeeman field. In the complex parameter
space of SC interaction, RSOC, Zeeman field and temperature we showed that thermal fluctuations
alter the FS topology and shifts the corresponding Dirac point from ${\bf k}=0$ to 
${\bf k} \neq 0$. Further, we established that a temperature controlled
FS segmentation is realizable in these systems, such that only parts of the FS serves as hotspots
for quasiparticle scattering. The response of the SC state to the Zeeman field is analyzed
based of the spectroscopic and thermodynamic signatures and the regimes of different SC pairings
are mapped out. The results presented in this letter provide benchmarks for 
the thermal scales and regime of stability of these systems against thermal fluctuations,
which are important from the perspective of spintronic devices. Our results
are compared with the experimental observations on Bi$_{2}$Te$_{3}$/NbSe$_{2}$ hybrid and are found
to be in remarkable qualitative agreement. It is expected that the results discussed
in this letter will trigger important experiments to capture the temperature controlled properties
of NCS and their response to in-plane Zeeman field.  A generic theoretical framework for finite
momentum scattering of quasiparticles and the associated FS signatures is provided, which should
be applicable to a wide class of SC materials. 

\textit{Acknowledgement:} The author acknowledges the use of the high performance computing 
cluster facility (AQUA) at IIT Madras, India. Funding from the Center for Quantum Information
Theory in Matter and Spacetime, IIT Madras is acknowledged. The author would like to thank
Avijit Misra for the critical reading of the manuscript. 

\bibliography{soc.bib}

\pagebreak
\widetext
\begin{center}
\textbf{\large Supplementary Material:Temperature tuned Fermi surface topology and segmentation in non-centrosymmetric superconductors}
\end{center}
\setcounter{equation}{0}
\setcounter{figure}{0}
\setcounter{table}{0}
\setcounter{page}{1}
\makeatletter
\renewcommand{\theequation}{S\arabic{equation}}
\renewcommand{\thefigure}{S\arabic{figure}}

\section{Method and indicators}

\subsection{Path integral formalism and Static Path Approximation (SPA)}
The attractive Hubbard model on a 2D square lattice with RSOC and in-plane Zeeman field along the
$x$-axis reads as,
\begin{eqnarray}
H & = & -\sum_{\langle ij\rangle, \sigma}t_{ij}(c_{i,\sigma}^{\dagger}c_{j,\sigma} + h. c.) - \mu\sum_{i,\sigma}\hat n_{i,\sigma} + \lambda\sum_{ij,\sigma,\sigma^{\prime}}(c_{i,\sigma}^{\dagger}(i\hat \sigma_{y})_{\sigma,\sigma^{\prime}}c_{j,\sigma^{\prime}} + c_{i,\sigma}^{\dagger}(-i\hat\sigma_{x})_{\sigma,\sigma^{\prime}}c_{j,\sigma^{\prime}}) \nonumber \\ && + h\sum_{i}(c_{i,\uparrow}^{\dagger}c_{i,\downarrow}+c_{i,\downarrow}^{\dagger}c_{i,\uparrow}) - \vert U \vert\sum_{i}\hat n_{i,\uparrow}\hat n_{i,\downarrow}  
\end{eqnarray}  
where, $t_{ij}=t=1$ is the amplitude of nearest neighbor hopping and sets the energy scale of the problem.
$\lambda$ is the strength of RSOC and $\mu$ is the global chemical potential. $\vert U\vert > 0$ is the onsite
Hubbard interaction. In terms of the Grassmann fields $\psi_{i,\sigma}(\tau)$ and
$\bar \psi_{i,\sigma}(\tau)$ we write the Hubbard partition function as,
\begin{eqnarray}
Z & = & \int {\cal D}\psi {\cal D}\bar \psi e^{-S[\psi, \bar \psi]}  
\end{eqnarray}
where, 
\begin{eqnarray}
S &=& \int_{0}^{\beta}d{\tau}[\sum_{ij, \sigma, \sigma^{\prime}}\{\bar \psi_{i,\sigma}((\partial_{\tau}-\mu)\delta_{ij}-t_{ij})\psi_{j,\sigma}\} + \lambda\sum_{ij,\sigma,\sigma^{\prime}}(\bar \psi_{i,\sigma}(i\hat \sigma_{y})_{\sigma,\sigma^{\prime}}\psi_{j,\sigma^{\prime}}+\bar\psi_{i,\sigma}(-i\hat \sigma_{x})_{\sigma,\sigma^{\prime}}\psi_{j,\sigma^{\prime}}) \nonumber \\ && - \vert U\vert \sum_{i, \sigma,\sigma^{\prime}}\bar \psi_{i,\sigma}\psi_{i,\sigma}\bar \psi_{i,\sigma^{\prime}}\psi_{i,\sigma^{\prime}} + h\sum_{i, \sigma, \sigma^{\prime}}\bar \psi_{i,\sigma}(\hat \sigma_{x})_{\sigma, \sigma^{\prime}}\psi_{i,\sigma^{\prime}}] \nonumber \\ &&   
\end{eqnarray}  
The interaction generates a quartic term in $\psi$ which can not be readily evaluated. In order to make
the model numerically tractable we decouple the quartic interaction exactly using Hubbard Stratonovich
decomposition \cite{hs1,hs2}. The decomposition introduces the complex scalar bosonic auxiliary fields $\Delta_{i}(\tau)$ and
$\Delta_{i}^{*}(\tau)$ which couples to the SC pairing. The corresponding partition function reads as,
\begin{eqnarray}
Z & = & \int {\cal D}\Delta {\cal D}\Delta^{*}{\cal D}\psi{\cal D}\bar\psi e^{-S_{1}[\psi, \bar \psi, \Delta, \Delta^{*}]} 
\end{eqnarray}  
where, the action is defined as,
\begin{eqnarray}
  S_{1} & = & \int_{0}^{\beta}d{\tau}[\sum_{ij, \sigma, \sigma^{\prime}}\{\bar \psi_{i,\sigma}((\partial_{\tau}-\mu)\delta_{ij}t_{ij})\psi_{j,\sigma}\} + \lambda\sum_{ij,\sigma,\sigma^{\prime}}(\bar \psi_{i,\sigma}(i\hat \sigma_{y})_{\sigma,\sigma^{\prime}}\psi_{j,\sigma^{\prime}}+\bar\psi_{i,\sigma}(-i\hat \sigma_{x})_{\sigma,\sigma^{\prime}}\psi_{j,\sigma^{\prime}}) \nonumber \\ && + \sum_{i}(\Delta_{i}(\tau)\bar \psi_{i,\uparrow}\bar \psi_{i,\downarrow} + \Delta_{i}^{*}(\tau)\psi_{i,\downarrow}\psi_{i,\uparrow} + \frac{\vert \Delta_{i}(\tau)\vert^{2}}{\vert U\vert}) 
  + h\sum_{i, \sigma, \sigma^{\prime}}\bar\psi_{i, \sigma}(\hat \sigma_{x})_{\sigma, \sigma^{\prime}}\psi_{i, \sigma^{\prime}}]  
\end{eqnarray}  
The $\psi$ integral is now quadratic but at the cost of additional integration over $\Delta_{i}(\tau)$ and
$\Delta_{i}^{*}(\tau)$. The weight factor for the $\Delta_{i}$ configurations can be determined by integrating
out the $\psi$ and $\bar \psi$ and using these weighted configurations one goes back and computes the fermionic
properties. Formally,

\begin{eqnarray}
Z &= & \int {\cal D}\Delta {\cal D}\Delta^{*}e^{-S_{2}[\Delta, \Delta^{*}]}  
\end{eqnarray}
where,
\begin{eqnarray}
S_{2} & = & {\mathrm {ln}}[{\mathrm {Det}}[{\cal G}^{-1}-\Delta_{i}(\tau)]] + \frac{\vert \Delta_{i}(\tau)\vert^{2}}{\vert U\vert}   
\end{eqnarray}
where, ${\cal G}$ is the electron Greens function in the $\{\Delta_{i}\}$ background.
The weight factor for an arbitrary space-time configuration $\Delta_{i}(\tau)$ involves the determination
of the fermionic determinant in that background.

In terms of the Matsubara modes we can write the bosonic
auxiliary field as $\Delta_{i}(\Omega_{n})$.
Within the purview of SPA we retain
only the $\Omega_{n}=0$ modes of the auxiliary fields, such that $\Delta_{i}(\Omega_{n}) \rightarrow \Delta_{i}$. 
Thus, we take into account the spatial fluctuations in $\Delta_{i}$ at all orders but retain only the
zero mode in the Matsubara frequency. The SPA reduces to BdG-MFT at $T = 0$,  while at $T\neq 0$ we retain
not just the saddle point configuration but ``all configurations'' based on the weight factor $e^{-S_{2}}$
above. Inclusion of the classical fluctuations on amplitude and phase suppresses the SC order quicker than
the MFT and allows the determination of the thermal scales with quantitative accuracy \cite{mpk2016,mpk2018,mpk_jpcm2020,mpk_lieb2020}.
%%%%%%%%%%%%%%%%%%%%%%%%%%%%%%%%%%%%%%%%%%%%%%%%%%%%%%%%%%%%%%%%%%%%%%%%%%%%%%%%%%%%%%%%%%
\begin{figure*}
\begin{center}
\includegraphics[height=12cm,width=15cm,angle=0]{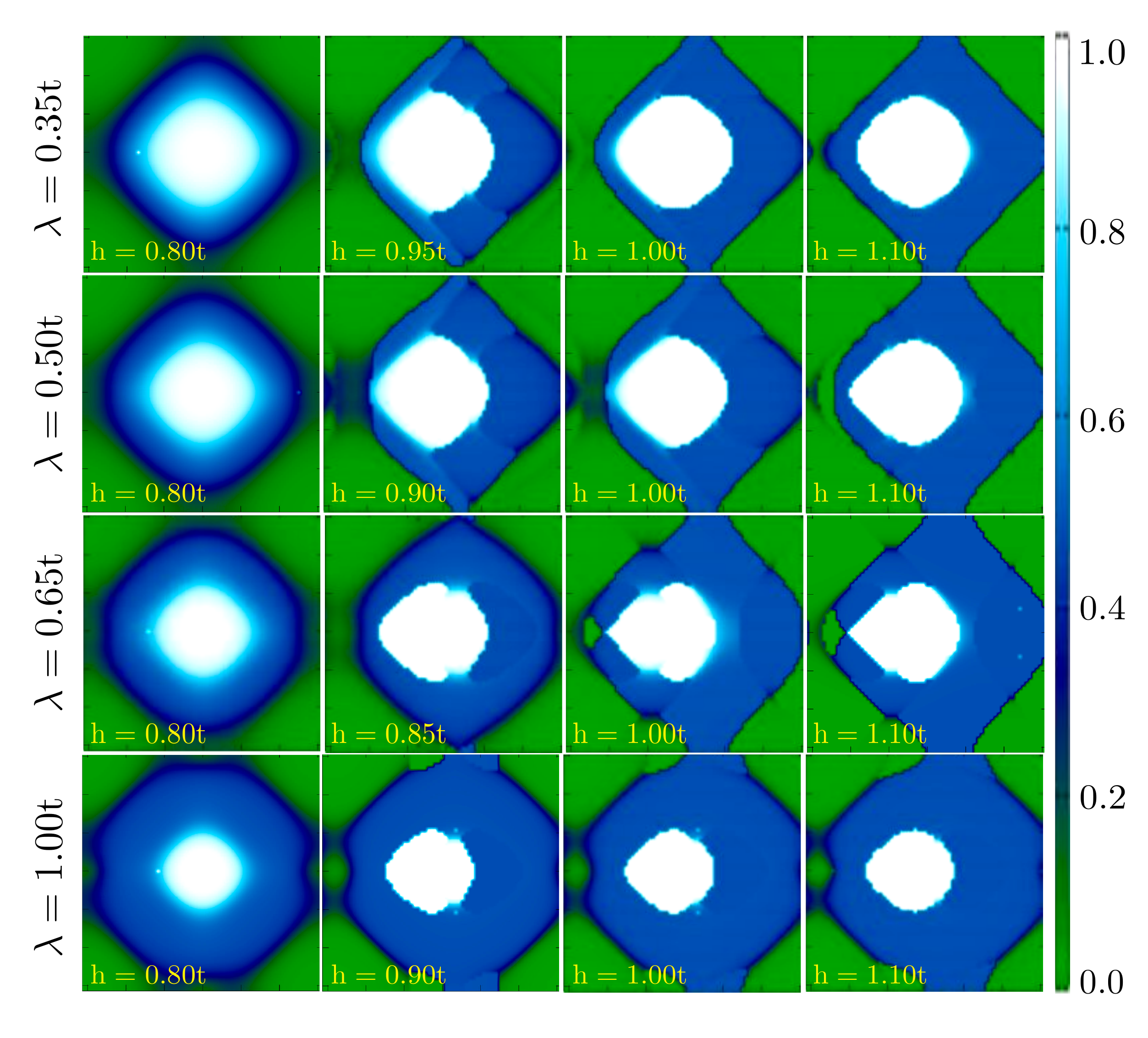}    
\caption{Evolution of the Fermi surface with in-plane Zeeman field ($h$) for 
selected RSOC ($\lambda$). Note the topological transition (TT) at intermediate
$\lambda-h$ cross sections, wherein a single self intersecting FS is realized. Zeeman effect
dominates at strong $\lambda$, with significant mismatch in the size of the FSs.
The calculations are carried out using BdG-MFT for a system size of $L=70$.}
\label{fig1s}
\end{center}    
\end{figure*}
%%%%%%%%%%%%%%%%%%%%%%%%%%%%%%%%%%%%%%%%%%%%%%%%%%%%%%%%%%%%%%%%%%%%%%%%%%%%%%%%%%%%%%%%%%%

\subsection{Monte Carlo simulation}
The random background configurations of $\{\Delta_{i}\}$ are generated numerically via Monte Carlo simulation
and obeys the Boltzmann distribution,
\begin{eqnarray}
P\{\Delta_{i}\} \propto Tr_{c,c^{\dagger}}e^{-\beta H_{eff}}  
\end{eqnarray}
where, the effective Hamiltonian $H_{eff}$ reads as,
\begin{eqnarray}
  H_{eff} &=& -t\sum_{\langle ij\rangle, \sigma}(c_{i,\sigma}^{\dagger}c_{j, \sigma} + H. c.) -\mu \sum_{i, \sigma}\hat n_{i, \sigma} +\lambda\sum_{ij, \sigma, \sigma^{\prime}}(c_{i, \sigma}^{\dagger}(i\hat \sigma_{y})_{\sigma, \sigma^{\prime}}c_{j, \sigma^{\prime}} + c_{i, \sigma}^{\dagger}(-i\hat \sigma_{x})_{\sigma, \sigma^{\prime}}c_{j, \sigma^{\prime}}) \nonumber \\ && + \sum_{i}(\Delta_{i}c_{i,\uparrow}^{\dagger}c_{i,\downarrow}^{\dagger} +\Delta_{i}^{*}c_{i,\downarrow}c_{i, \uparrow}) + \sum_{i}\frac{\vert \Delta_{i}\vert^{2}}{\vert U \vert} 
  + h\sum_{i,\sigma, \sigma^{\prime}}c_{i,\sigma}^{\dagger}(\hat \sigma_{x})_{\sigma,\sigma^{\prime}}c_{i,\sigma^{\prime}}  
\end{eqnarray}  
For large and random configurations the trace is computed numerically, wherein $H_{eff}$ is diagonalized for
each attempted update of $\Delta_{i}$ and converge to the equilibrium configuration via Metropolis algorithm.
The process is numerically expensive and involves a computation cost of ${\cal O}(N^{3})$ per update (where
$N=L\times L$ is the number of lattice sites), thus
the cost per MC sweep is $\sim N^{4}$. The computation cost is cut down by using a traveling cluster
algorithm (TCA), wherein instead of diagonalizing the entire Hamiltonian for each attempted update of
${\Delta_{i}}$, we diagonalize a smaller cluster of size $N_{c}\times N_{c}$ surrounding the update site.
The corresponding computation cost now scales as ${\cal O}(NN_{c}^{3})$ which is linear in $N$. This allows
us to access larger system size with reasonable computation cost. The equilibrium configurations obtained
via the combination of MC and Metropolis at different temperatures are used to determine the fermionic
correlators \cite{mpk2016,mpk2018,mpk_jpcm2020,mpk_lieb2020}. 

\subsection{Bogoliubov-de-Gennes mean field theory}
We have used BdG-MFT as an alternate scheme to calculate the ground state properties of the system.
The use of MFT allows us to access large system sizes and our
ground state calculations are carried out on a system size of $L=70$. The free energy
minimization scheme involves optimization over trial solutions w. r. t. $\vert \Delta_{i}\vert$
and ${\bf q}$, where $\vert \Delta_{i}\vert$ is real and $\vert \Delta_{\bf q} \vert = \Delta_{0}$
corresponds to the amplitude of the uniform SC state. For the FFLO phase we choose different trial
solutions corresponding to the, 
(i) uniaxial modulation $\Delta_{i} \propto \Delta_{0}\cos(qx_{i})$, (ii) 2D modulation
$\Delta_{i} \propto \Delta_{0}(\cos(qx_{i})+\cos(qy_{i}))$ and (iii) diagonal modulation
$\Delta_{i} \propto \Delta_{0}\cos[q(x_{i}+y_{i})]$. For the helical SC phase the trial
solution is defined as, $\Delta_{i} \propto \Delta_{0} e^{i{\bf q.r}}$.
We work in the grand canonical ensemble and for $\mu \in [0:-4]$ with $\delta\mu = 0.5$
the free energy optimization is carried out for different $\lambda$ and $h \in [0:1.5]$,  in each case.
Irrespective of the choice of $\lambda$ the low magnetic field regime $h\le h_{c1}$ hosts an
uniform SC phase with ${\bf q}=0$. Over the regime $h_{c1} < h \le h_{c2}$ a stable helical
SC phase is realized, except close to $\mu \sim 0$, where a FFLO phase is stabilized for certain
choices of $\lambda$. The
$\mu$ dependence of $h_{c1}$ and $h_{c2}$ is weakly non monotonic, with an increase in the
regime $0 < \mu \le -t$, followed by a monotonic reduction for $-t < \mu \le -4t$. The
choice of $\mu=-t$ for the results presented in the main text is dictated by the maximum
$h_{c2}$ at this chemical potential, which allows a wider regime of helical SC state.

\subsection{Indicators}
The various phases of the system are characterized based on the following indicators,
\begin{itemize}
\item[(1)]{Pairing field structure factor,
\begin{eqnarray}
S({\bf q}) &=& \frac{1}{N^{2}}\sum_{ij}\langle \Delta_{i}\Delta_{j}^{*}\rangle e^{i{\bf q}.({\bf r}_{i}-{\bf r}_{j})}    
\end{eqnarray}    
where, $N=L\times L$ is the number of lattice sites.
}
\item[(2)]{Single particle density of states (DOS),
\begin{eqnarray}
N(\omega) &=& \frac{1}{N}\langle \sum_{i,n}(\vert u_{n}^{i}\vert^{2}\delta(\omega-E_{n})+\vert v_{n}^{i}\vert^{2}\delta(\omega + E_{n}))\rangle    
\end{eqnarray}
where, $u_{n}^{i}$ and $v_{n}^{i}$ are the Bogoliubov-de-Gennes (BdG) eigenvectors corresponding to the eigenvalue $E_{n}$. 
}
\item[(3)]{Spectral function,
\begin{eqnarray}
A({\bf k}, \omega) & = & -(1/\pi)\Im G({\bf k}, \omega)
\end{eqnarray}
here, $G({\bf k}, \omega)=lim_{\delta \rightarrow 0}G({\bf k}, i\omega_{n})\vert_{i\omega_{n}\rightarrow \omega + i\delta}$, with, $G({\bf k}, i\omega_{n})$ being the imaginary frequency transform of $\langle c_{\bf k}(\tau)c_{\bf k}^{\dagger}(0)\rangle$. 
}
\item[(4)]{Low energy spectral weight distribution,
\begin{eqnarray}
A({\bf k}, 0) & = & -(-1/\pi)\Im G({\bf k}, 0)
\end{eqnarray}}
\item[(5)]{Momentum resolved fermionic occupation number,
\begin{eqnarray}
n({\bf k}) & = & \sum_{\sigma}\langle c_{{\bf k}, \sigma}^{\dagger}c_{{\bf k}, \sigma}\rangle 
\end{eqnarray}}  
\end{itemize}  

\section{Effect of SOC:- Fermi surface evolution}
We show the RSOC dependence of FS topology in Fig.\ref{fig1s}. For $h > h_{c1}$ a helical SC state
with finite-$q$ pairing is realized for weak and intermediate RSOC. For strong RSOC the
Zeeman effect dominates and there is significant mismatch in the size of the helical FSs;
superconductivity is strongly suppressed in this regime. The topological transition (TT)
between the inter and intraband helical SC pairing is most prominent over the regime of intermediate
RSOC ($\lambda=0.50t$ and $\lambda=0.65t$). The TT can thus be controlled by suitably tuning the
combination of RSOC strength and the in-plane Zeeman field.
%%%%%%%%%%%%%%%%%%%%%%%%%%%%%%%%%%%%%%%%%%%%%%%%%%%%%%%%%%%%%%%%%%%%%%%%%%%%%%%%%%%%%%%%%%%
\begin{figure*}
\begin{center}
\includegraphics[height=7.0cm,width=14.5cm,angle=0]{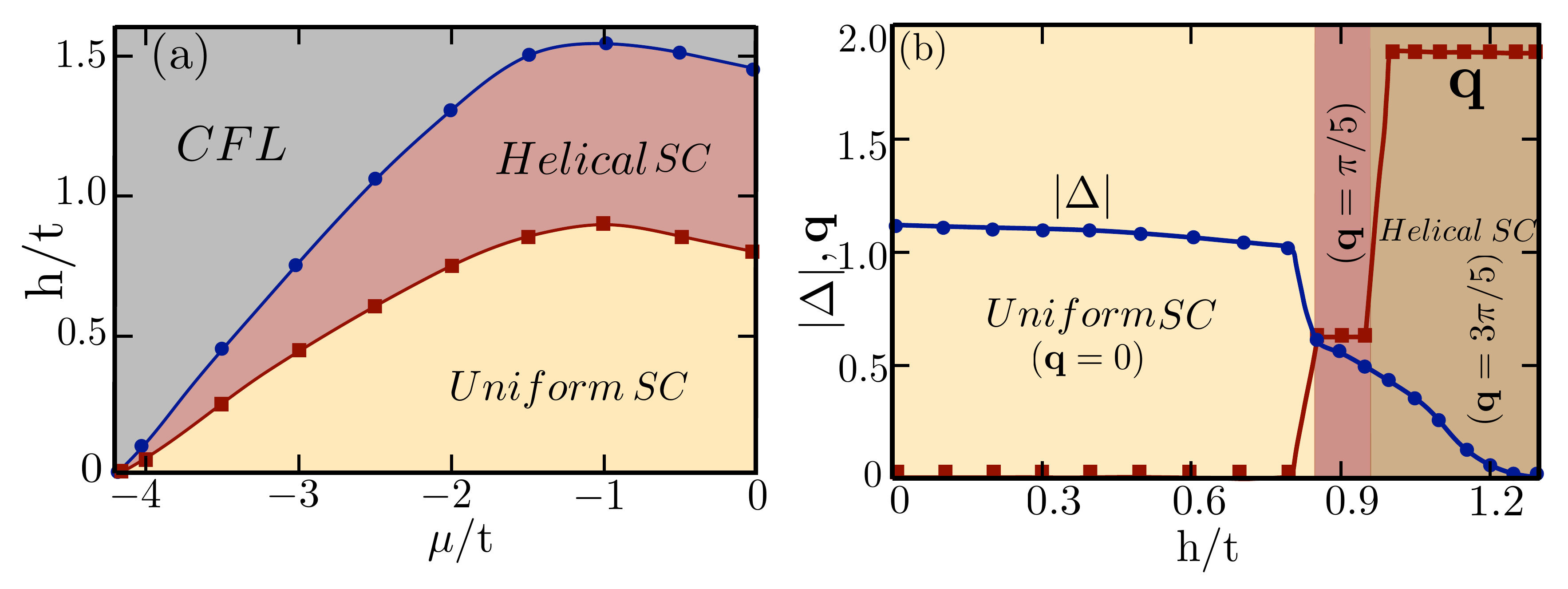}
\caption{(a) Ground state phase diagram in the $\mu-h$ plane for the selected RSOC of $\lambda=0.65t$.
  The thermodynamic phases include uniform and helical SC phases with the corresponding critical
  fields $h_{c1}$ and $h_{c2}$, respectively. (b) Zeeman field dependence of average superconducting
  pairing field amplitude ($\vert \Delta \vert$) and pairing momentum (${\bf q}$), at $\lambda=0.65t$,
  in the ground state. The first order transition between the uniform and helical SC phases is signaled
  by the sharp discontinuity in $\vert \Delta\vert$ and ${\bf q}$. The helical phase comprises of uniaxial
  modulations with the pairing momenta being $(0, \pi/5)$ and $(0, 3\pi/5)$.}
\label{fig2s}
\end{center}    
\end{figure*}
%%%%%%%%%%%%%%%%%%%%%%%%%%%%%%%%%%%%%%%%%%%%%%%%%%%%%%%%%%%%%%%%%%%%%%%%%%%%%%%%%%%%%%%%%%%

\section{Ground state}
We map out the ground state phase diagram of the system in the $\mu-h$ plane at $\lambda=0.65t$,
in Fig.\ref{fig2s}(a). The regimes of ${\bf q}=0$ and ${\bf q} \neq 0$ pairing are demarcated
as the uniform and helical SC states, respectively. RSOC stabilizes a helical SC state over the
entire $\mu$ regime, unlike the situation in the absence of RSOC where application of an
in-plane Zeeman field gives rise to an amplitude modulated FFLO state \cite{mpk2016}.  
For the rest of our calculations we have fixed $\mu = -t$, corresponding to an electron number
density of $n \approx 0.75t$. 

In Fig.\ref{fig2s}(b) we show the behavior of the mean pairing field amplitude
$\vert \Delta\vert = \langle\vert \Delta_{i}\vert \rangle$ and the pairing momentum ${\bf q}$,
at $\mu=-t$ and $\lambda=0.65t$. Over the uniform SC regime
($0 < h \le h_{c1}$) the $\vert \Delta\vert$ is nearly constant and the corresponding pairing
momentum is at ${\bf q}=0$. At $h=h_{c1}$ the system undergoes a first order transition to
the helical SC state, accompanied by a strong suppression in the pairing field amplitude.
Concomitantly, the pairing momentum becomes finite with ${\bf q} \neq 0$. While $\vert \Delta\vert$
undergoes continuous suppression with $h$, the pairing momenta remains largely constant over
the regime $h_{c1} < h \le h_{tp}$. The topological transition between the inter and intraband
helical SC states at $h_{tp}$ is accompanied by yet another first order transition, as suggested
by the discontinuity in the pairing momenta ${\bf q}$. The jump discontinuity of ${\bf q}$ across
the topological transition to a single self intersecting FS has been recently discussed within the
purview of MFT \cite{fu_pnas2021}.  
Note that this topological transition can not be detected via real space signatures and standard
experimental probes quantifying the SC transitions.
%%%%%%%%%%%%%%%%%%%%%%%%%%%%%%%%%%%%%%%%%%%%%%%%%%%%%%%%%%%%%%%%%%%%%%%%%%%%%%%%%%%%%%%%%%%
\begin{figure*}
\begin{center}
\includegraphics[height=10.5cm,width=14.5cm,angle=0]{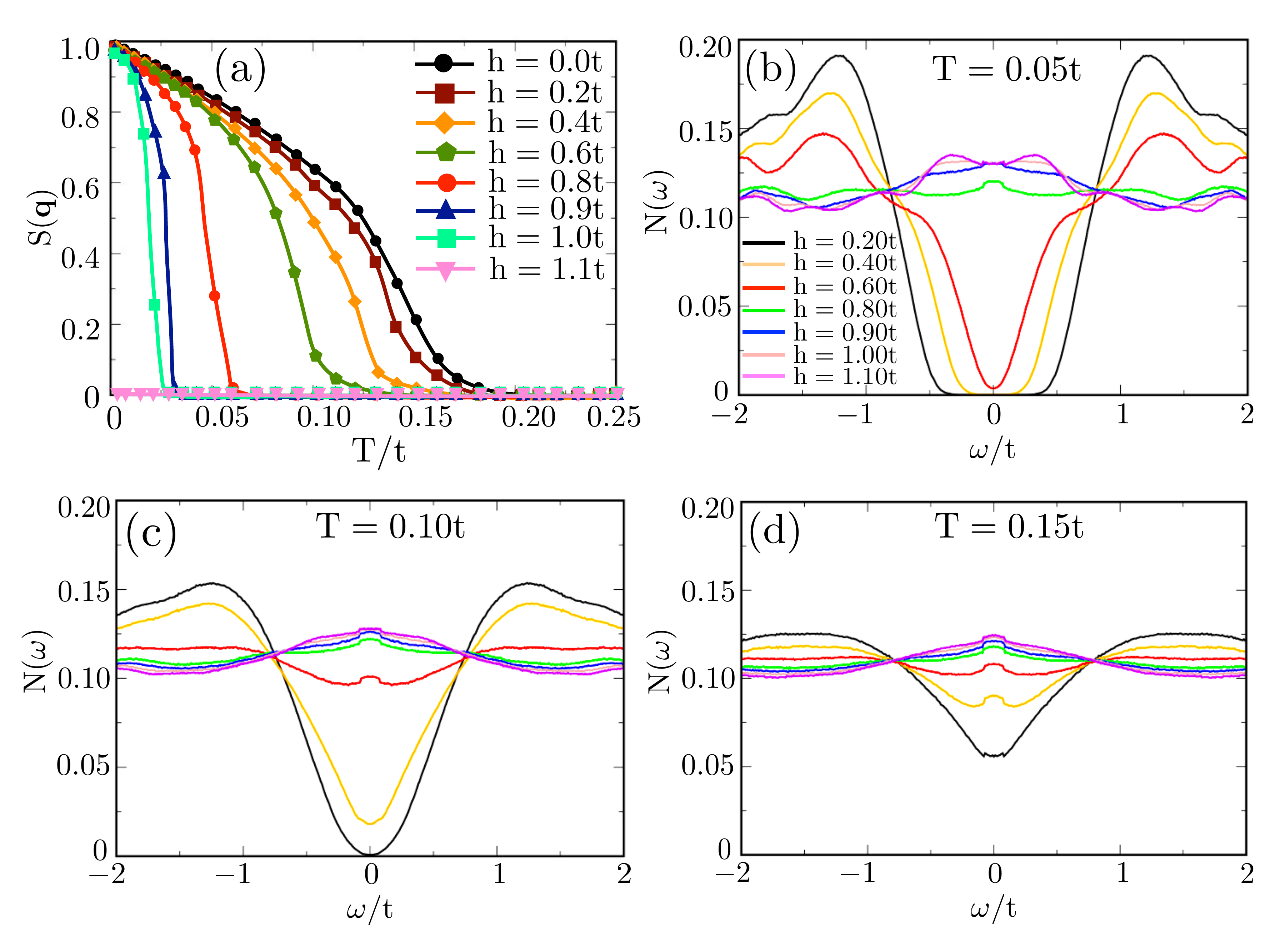}
\caption{(a) Temperature dependence of the pairing field structure factor
  ($S({\bf q})$) at $\lambda=0.65t$ and different Zeeman field. The point of inflection of each curve corresponds
  to the respective $T_{c}$. (b)-(c) Zeeman field dependence of the single particle DOS at the Fermi level
  for the selected temperatures of $T=0.05t$, $T=0.10t$ and $T=0.15t$, respectively.}
\label{fig3s}
\end{center}    
\end{figure*}
%%%%%%%%%%%%%%%%%%%%%%%%%%%%%%%%%%%%%%%%%%%%%%%%%%%%%%%%%%%%%%%%%%%%%%%%%%%%%%%%%%%%%%%%%%%

\section{Determination of the thermal scales}
The transition temperature ($T_{c}$) corresponding to the loss of (quasi) long range SC phase
coherence, shown in the main text, is determined based on the temperature dependence of the
pairing field structure factor ($S({\bf q})$). We show the same in Fig.\ref{fig3s}(a) as
function of the Zeeman field. The point of inflection of each curve correspond to the respective
$T_{c}$. The Zeeman field strongly suppresses the $T_{c}$ and at $h \sim h_{c1}$ a first
order transition between the uniform and the helical SC phases is realized. Within the helical SC
phase ($h_{c1} < h \le h_{c2}$) the $S({\bf q})$ changes with the Zeeman field via consecutive
first order transitions. The system loses SC order at $h\sim h_{c2}$, as signaled by the vanishing $S({\bf q})$.

In Fig.\ref{fig3s}(b)-(d) we show the field dependence of the single particle DOS at the Fermi
level at three different thermal cross sections. At $T=0.05t$ (Fig.\ref{fig3s}(b)) the system
is a gapped uniform superconductor at $h=0.2t$ with a robust spectral gap at the Fermi level. Increasing
$h$ suppresses the gap progressively as observed at $h=0.4t$ and finally leads to its closure
at $h=0.6t$, indicated by a finite spectral weight at the Fermi level. $h=0.6t$ corresponds to
a gapless SC state (see Fig.1(a) in the main text) with (quasi) long range SC phase coherence
indicated by the broadened but prominent QP peaks at the gap edges. The system undergoes transition
to a regime with helical SC correlations at $h \ge 0.9t$, characterized by a gapless spectra and
thermally diffused signatures of in-gap states. Zeeman field progressively suppresses the in-gap
states and at $h \sim 1.1t$ the spectra is featureless, corresponding to a magnetic metal. 

At $T=0.10t$ (Fig.\ref{fig3s}(c))
the system is a uniform SC with suppressed spectral gap at $h=0.2t$ and a gapless superconductor
at $h=0.4t$. A change in the FS topology with the shift in the Dirac point to ${\bf k} \neq 0$
takes place for $h=0.6t$ at $T=0.10t$ (see Fig.1(a) of the main text). The corresponding single
particle DOS shows a sudden large spectral
weight accumulation at the Fermi level which we attribute to the short range helical SC correlations.
The coherence peaks at the gap edges are strongly suppressed via transfer of spectral weight away
from the Fermi level. Further
increase in $h$ progressively accumulates the spectral weight at the Fermi level as the system
crosses over to the pseudogap phase and eventually to the CFL. The $T=0.15t$ (Fig.\ref{fig3s}(d))
cross section corresponds to a gapless superconductor at low $h$, which progressively gives way
to the psuedogap and then to the CFL phases via spectral weight accumulation at the Fermi level.

\end{document}